\documentclass[aps,prl,letterpaper,twocolumn,floatfix,superscriptaddress]{revtex4-2}
\usepackage{amsmath,amssymb} 
\usepackage[T1]{fontenc}
\usepackage[utf8]{inputenc} 

\usepackage{graphicx} 

\usepackage{hyperref}
\hypersetup{colorlinks=true,citecolor={black},linkcolor={black},urlcolor={black}} 
\usepackage{xcolor}
\usepackage{bm}

\graphicspath{{.}{./figs/}}

\begin{document}

\title{Neural Networks with Exciton-Polariton Waveguides}

\affiliation{Beijing Academy of Quantum Information Sciences, Beijing, 100193, China}
\affiliation{State Key Laboratory of Low-Dimensional Quantum Physics and Department of Physics, Tsinghua University, Beijing, 100084, China}
\affiliation{Frontier Science Center for Quantum Information, Beijing, 100084, China}
\affiliation{Collaborative Innovation Center of Quantum Matter, Beijing, 100084, China}
\affiliation{Laboratory of Materials Modification by Laser, Ion and Electron Beams, Ministry of Education, Dalian University of Technology, Dalian, 116024 China}
\affiliation{Institute of Physics, Polish Academy of Sciences, Al. Lótnikow 32/46, Warsaw PL-02-668, Poland}
\affiliation{Division of Physics and Applied Physics, School of Physical and Mathematical Sciences, Nanyang Technological University, 21 Nanyang Link, Singapore 637371, Singapore}
\affiliation{MajuLab, International Joint Research Unit UMI 3654, CNRS, Universit\'e C\^ote d'Azur, Sorbonne Universit\'e, National University of Singapore, Nanyang Technological University, Singapore}

\author{Huawen Xu}
\email{xuhw@baqis.ac.cn}
\affiliation{Beijing Academy of Quantum Information Sciences, Beijing, 100193, China}
\affiliation{Huawen Xu and Ruiqi Bao contributed equally to this work.}

\author{Ruiqi Bao}
\affiliation{Laboratory of Materials Modification by Laser, Ion and Electron Beams, Ministry of Education, Dalian University of Technology, Dalian, 116024 China}
\affiliation{Huawen Xu and Ruiqi Bao contributed equally to this work.}

\author{Lingyu Tian}
\affiliation{Beijing Academy of Quantum Information Sciences, Beijing, 100193, China}

\author{Yusong Gan}
\affiliation{State Key Laboratory of Low-Dimensional Quantum Physics and Department of Physics, Tsinghua University, Beijing, 100084, China}

\author{Junfeng Gao}
\affiliation{Laboratory of Materials Modification by Laser, Ion and Electron Beams, Ministry of Education, Dalian University of Technology, Dalian, 116024 China}

\author{Michal Matuszewski}
\affiliation{Institute of Physics, Polish Academy of Sciences, Al. Lótnikow 32/46, Warsaw PL-02-668, Poland}

\author{Qihua Xiong}
\email{qihua_xiong@tsinghua.edu.cn}
\affiliation{State Key Laboratory of Low-Dimensional Quantum Physics and Department of Physics, Tsinghua University, Beijing, 100084, China}
\affiliation{Beijing Academy of Quantum Information Sciences, Beijing, 100193, China}
\affiliation{Frontier Science Center for Quantum Information, Beijing, 100084, China}
\affiliation{Collaborative Innovation Center of Quantum Matter, Beijing, 100084, China}

\author{Timothy C. H. Liew}
\email{tchliew@gmail.com}
\affiliation{Division of Physics and Applied Physics, School of Physical and Mathematical Sciences, Nanyang Technological University, 21 Nanyang Link, Singapore 637371, Singapore}
\affiliation{MajuLab, International Joint Research Unit UMI 3654, CNRS, Universit\'e C\^ote d'Azur, Sorbonne Universit\'e, National University of Singapore, Nanyang Technological University, Singapore}

\begin{abstract}

We propose and numerically study a scheme for neural networks using exciton-polariton waveguide meshes. These enable arbitrary programmable interlayer couplings suited to a restricted Boltzmann machine architecture where polariton phases act as stochastic network variables. We show that such a network can be trained using unsupervised Hebbian and anti-Hebbian updates. This allows the implementation of an associative memory, where the network recovers learned patterns from imperfect inputs.

\end{abstract}

%\date{\today}

\maketitle

%%%%%%%
%\textit{Introduction---}
Exciton-polaritons are hybrid states of light and matter, renowned for borrowing concepts from other fields of research, while promising optoelectronic devices through the balance of different physical properties~\cite{Sanvitto2016}. For example, inspiration from biological neurons was used to design mechanisms of robust signal propagation~\cite{Liew2008}, suitable for universal polariton-based circuits~\cite{EspinosaOrtega2013}, while proposals of neural networks based on the interference of polariton fields~\cite{EspinosaOrtega2015} and reservoir computing~\cite{Opala2019} also emerged. The latter has been popular in experimental demonstrations~\cite{Ballarini2020,Gan2025,Long2025} as it replaces the need for precise control of inter-neuron coupling weights needed in most neural network architectures with the need for a larger set of randomly connected neurons. Ultimately though, this brings a cost of inefficient network sizes and, so far, all demonstrations have required at least a part of the network operation to be performed externally (typically a linear transformation of outputs or some other post-processing). Many polariton network architectures~\cite{Mirek2021,Matuszewski2024,Sedov2025}, including those considered for the design of spin Hamiltonian simulators~\cite{Kalinin2020}, also tend to suffer from a problem of limited connectivity. They have benefited from decades of research into polariton lattices~\cite{Amo2016,Schneider2017,Klembt2024}, however, as polariton lattices tend to be fully described by nearest neighbour interactions (or occasionally next nearest neighbour interactions) they do not reach the massive connectivity (all-to-all coupling) that brings the real power to neural network architectures. The theoretically desirable Hopfield form of a real symmetric coupling matrix with fully controllable elements $J_{ij}$ has been elusive.

Although lattices have been the mainstream of exciton-polariton physics, the propagation of polaritons in channels and waveguides has also frequently appeared as a foundation for device concepts. Early works showed the possibility of creating polariton interferometers~\cite{Sturm2013}, amplifiers~\cite{Wertz2012}, switches~\cite{Gao2012}, and couplers~\cite{Rozas2020}. Polariton channels are compatible with room temperature operation~\cite{Su2018}, and one can also work in the regime of total internal reflection, which allows the realization of high-quality factor systems, capable of polariton condensation (the formation of a coherent state) and electrically injected polariton lasing~\cite{Souissi2026}, with modest demands on fabrication processes. Beam-splitters~\cite{Fieramosca2025} and local phase shifters have been introduced~\cite{SuarezForero2021,Coriolano2025}, which can also be electrically controlled~\cite{Rosenberg2016,Liran2018}. In linear photonics, the Reck-Clements schemes~\cite{Reck1994,Clements2016} realize an arbitrary unitary operation using beam splitters and local phase shifters. A Reck-Clements mesh constructed from these polariton-based components can in principle implement a programmable unitary transformation $U_{ij}$. 

\begin{figure}[t]
\centering
\includegraphics[width=0.5\textwidth]{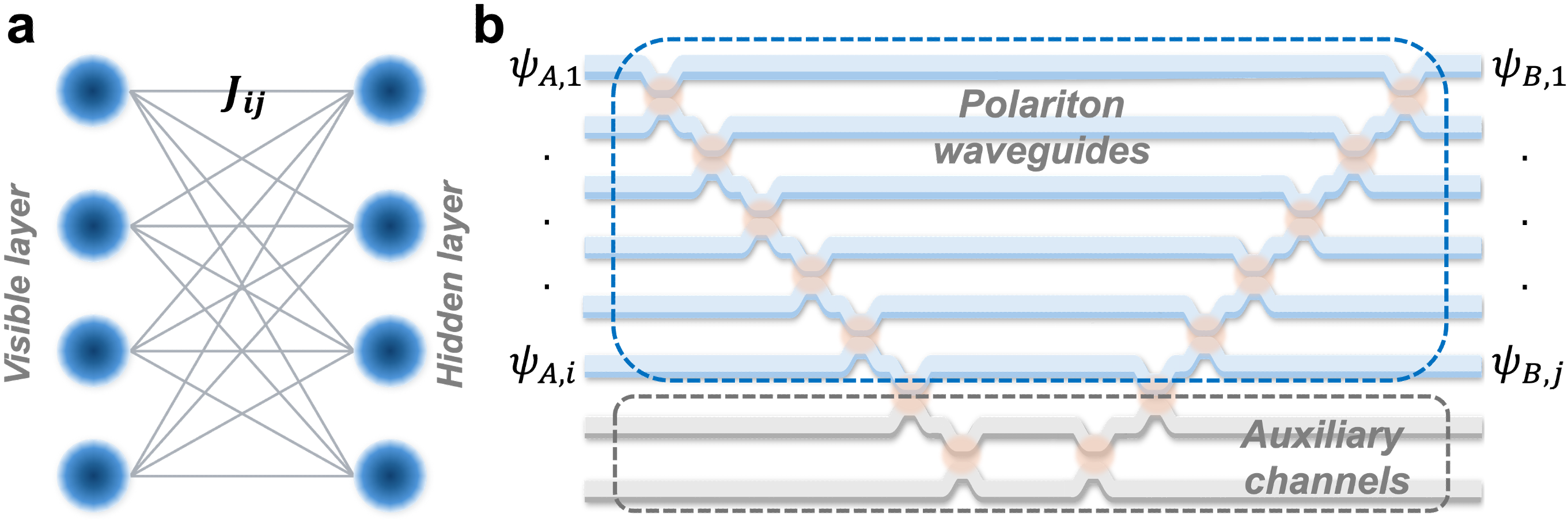}
\caption{
Schematic of the polariton-waveguide network. \textbf{a}, bipartite connectivity between the visible and hidden polariton nodes through the coupling matrix \(J\). \textbf{b}, the physical implementation in which the visible nodes on side \(A\) and the hidden nodes on side \(B\) are located at the two driven sides of the same reciprocal waveguide mesh. Auxiliary channels are used only to implement the desired inter-layer transformation and do not represent neural variables.
}
\label{Fig_schematic}
\end{figure}

In the present work we consider a neural network where the neural nodes are implemented with a set of waveguides, each capable of supporting a polariton condensate under non-resonant driving. The polaritons propagate through the aforementioned waveguide mesh. If this network is supplemented with some auxiliary waveguides that do not themselves input any signal, then the effect of the unitary transform on the original subset of waveguides can be shown to be equivalent to a real symmetric coupling matrix. At the ends of these waveguides, the fields are expected to be reflected. Application of non-resonant driving at their ends then reinforces the backward propagating condensates. In other words, we have a recurrent system, where polariton signals propagate back and forth and different waveguides are coupled by controllable weights $J_{ij}$, that is, of Hopfield form.

We propose that the recurrent bipartite architecture is effectively a restricted Boltzmann machine (RBM)~\cite{Byrnes2013}. Conventional RBMs use binary variables with sigmoid conditional probabilities. Here, the variables are polariton phases with a different probability distribution (which we will identify), and the long-range reciprocal coupling is implemented within the waveguide mesh rather than through external feedback. We will introduce a simple stochastic model of the system and demonstrate numerically its unsupervised learning and performance for an exemplary task.

%\textit{Model---}
A restricted Boltzmann machine is a bipartite neural network formed by a visible layer and a hidden layer, as illustrated in Fig.~\ref{Fig_schematic}\textbf{a}. In the conventional binary case, the visible and hidden variables, denoted by $v_i$ and $h_j$, take values $\pm1$ and are sampled from conditional probabilities of the form
\begin{align}
p\left(h_j=+1|v\right)&=f\left(\sum_{i}J_{ij}v_i\right),\notag\\
p\left(v_i=+1|h\right)&=f\left(\sum_{j}J_{ij}h_j\right),
\label{eq_RBM}
\end{align}
where $f$ is the activation function and $J_{ij}$ are the inter-layer weights. The essential structural feature is the absence of intra-layer connections: all couplings connect a visible degree of freedom to a hidden degree of freedom.

\begin{figure}[b]
\centering
\includegraphics[width=0.5\textwidth]{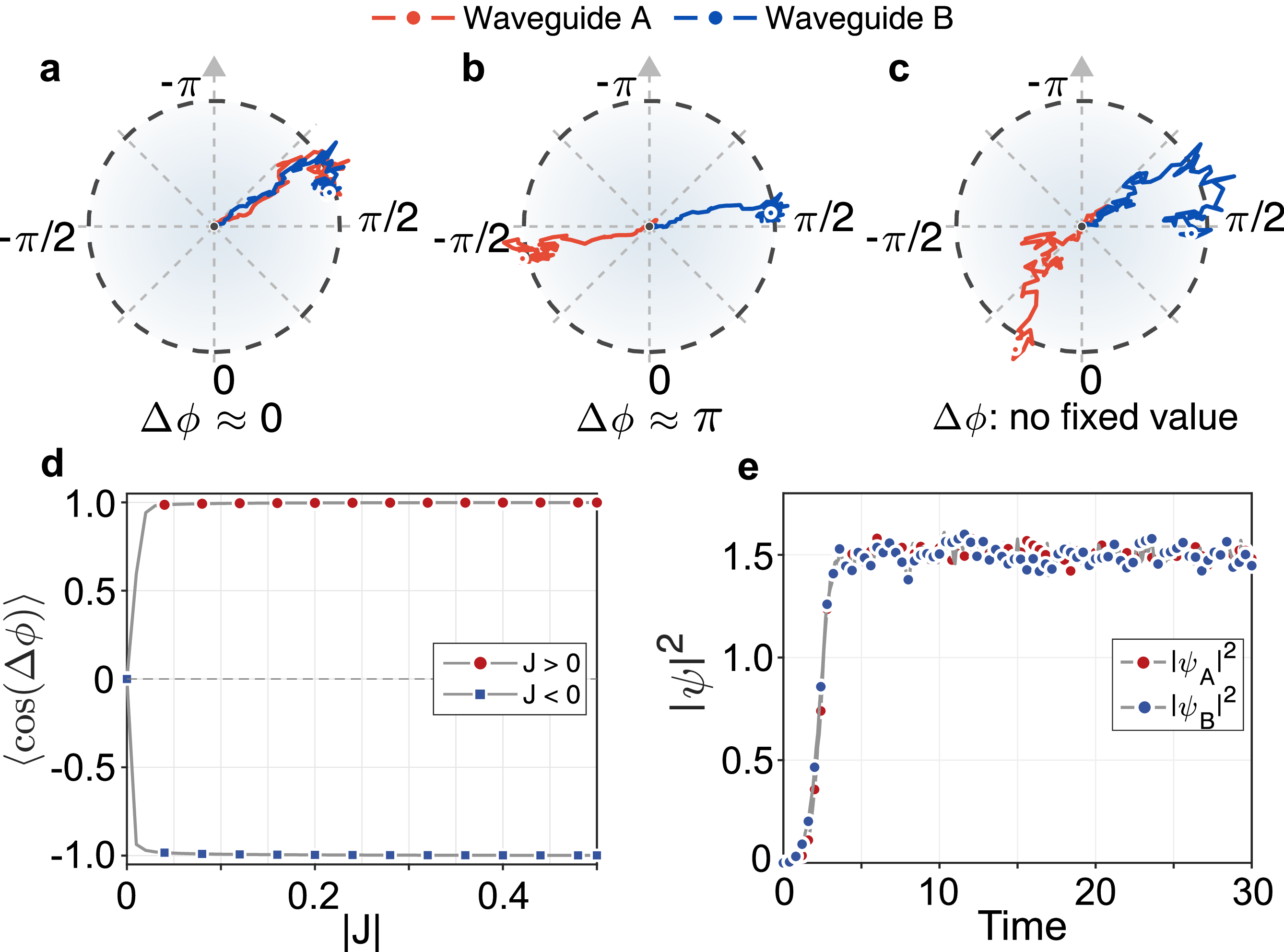}
\caption{
Phase locking in two coupled polariton waveguides.
\textbf{a--c}, Trajectories of the fields in the complex plane for
\(J=+1\), \(J=-1\), and \(J=0\), respectively. The polar angle represents
the phase of each field, while the radial coordinate represents the
normalized intensity. The angular separation of the two trajectories gives
the relative phase \(\Delta\phi\). Positive coupling produces in-phase
locking, negative coupling produces anti-phase locking, and the uncoupled
fields retain no fixed phase relation.
\textbf{d}, Signed phase correlation
\(\langle\cos(\Delta\phi)\rangle\) as a function of \(|J|\).
\textbf{e}, Intensity dynamics for \(J=+1\).
}
\label{Fig_dynamics}
\end{figure}

We implement this bipartite architecture using polariton condensate modes at the two driven sides of the same reciprocal waveguide network, shown in Fig.~\ref{Fig_schematic}\textbf{b}. The device contains $N$ signal waveguides. Non-resonant pumping (a local gain) at side $A$ supports the visible fields $\psi_{A,i}$, while pumping at side $B$ supports the "hidden" fields $\psi_{B,j}$. The visible and "hidden" variables are therefore the terminal polariton modes at the two driven sides of a single reciprocal device. The inter-layer weight matrix is replaced by a physical coupling matrix $J$, where the row index $i$ labels a node in the visible set $A$ and the column index $j$ labels a node in the hidden set $B$. 

Consider first the polariton fields on side $A$ of the network. Under non-resonant driving, each node obeys a stochastic gain-saturation equation,
\begin{equation}
d\psi_{A,i}
=
\left[
\left(P-1-|\psi_{A,i}|^2\right)\psi_{A,i}
+F_{A,i}
\right]dt
+\sqrt{2\sigma}\,d\eta_{A,i},
\label{eq:dpsiA}
\end{equation}
where $P$ is the pump strength, $F_{A,i}$ is a coherent input field (to be defined shortly), and $d\eta_{A,i}$ is an independent complex Wiener increment. We choose the temporal and intensity scales such that the linear and nonlinear loss rates are normalized to unity.

In the regime of weak amplitude fluctuations, the field amplitude remains close to $\sqrt{P-1}$ and the steady-state phase follows a von Mises distribution:
\begin{equation}
p(\phi_i|F_{A,i})\propto\exp[\kappa_i\cos(\phi_i-\phi_{F,i})],
\label{eq:Mises}
\end{equation}
where $F_{A,i}=|F_{A,i}|\exp(i\phi_{F,i})$ and $\kappa_i=|F_{A,i}|\sqrt{P-1}/\sigma$. The phase of the coherent input therefore sets the preferred polariton phase, while $\kappa_i$ controls the width of its stochastic response. The derivation, normalization, and range of validity of this phase distribution are given in the Supplemental Material (SM). Although the von Mises distribution is not a typical activation function, for the functioning of a restricted Boltzmann machine, it is enough that an activation function exists, in accordance with Eq.\ref{eq_RBM}. The fields $\psi_{A,i}$ are then fed into a programmable polariton waveguide mesh implementing a unitary transformation, following the Reck-Clements scheme. 

After an overall normalization, the real coupling matrix \(J\) is embedded as a subblock of a higher-dimensional unitary transformation using unexcited auxiliary channels, whose outputs are excluded from the recurrent dynamics. The construction of the enlarged unitary matrix is detailed in the SM. The fields driving the nodes on side $B$ are then written as $F_{B,j}=\sum_i J_{ij}\psi_{A,i}$, or $F_B=J^T\psi_A$. The nodes on side $B$ are also non-resonantly driven and evolve according to
\begin{equation}
d\psi_{B,j}
=
\left[
\left(P-1-|\psi_{B,j}|^2\right)\psi_{B,j}
+F_{B,j}
\right]dt
+\sqrt{2\sigma}\,d\eta_{B,j}.
\label{eq:dpsiB}
\end{equation}
After reflection, the fields pass back through the same network. This gives the driving fields on side $A$ as $F_{A,i}=\sum_j J_{ij}\psi_{B,j}$. We have neglected here the propagation time delay for the reflected fields. As time delay can nevertheless be important in polaritonics \cite{Topfer2020}, this assumption is lifted in the SM. Because the network contains no intra-layer couplings, the phase variables within one layer are conditionally independent when the opposite layer is fixed. Within the fixed-amplitude approximation used above, the single-mode von Mises responses together with the reciprocal couplings $J$ and $J^T$ define a bipartite continuous-phase Boltzmann distribution.

%\textit{Results---}
We first consider two coupled polariton nodes to determine how the sign of $J$ controls their relative phase. In a standard RBM, the entries of $J$ are statistical weights; here, they are coherent couplings between complex fields. Figures~\ref{Fig_dynamics}\textbf{a--c} show the two fields in the complex plane for \(J=+1\), \(J=-1\), and \(J=0\). The initial fields are taken as zero and evolve under a continuous pump $P$. Positive coupling brings the trajectories to the same angle, whereas negative coupling drives them to angles separated by \(\pi\). Without coupling, the angular separation continues to wander. The signed phase correlation in Fig.~\ref{Fig_dynamics}\textbf{d} distinguishes these two locked states: \(\langle\cos(\Delta\phi)\rangle\) tends toward \(+1\) for \(J>0\) and \(-1\) for \(J<0\) as \(|J|\) increases. The intensities remain noisy after locking, as illustrated for \(J=+1\) in Fig.~\ref{Fig_dynamics}\textbf{e}.

We next apply this phase-locking response to a multi-node associative-memory network. Each binary variable is encoded in the phase of the corresponding polariton field,
\begin{equation}
x_i=0\ \Rightarrow\ \phi_i=0,
\qquad
x_i=1\ \Rightarrow\ \phi_i=\pi.
\end{equation}
Equivalently, the field is initialized as
\begin{equation}
\psi_i=\sqrt{P-1}\exp(i\pi x_i).
\end{equation}
A stored pattern is a phase configuration distributed across the polariton modes.

The coupling matrix is trained by alternating Hebbian and anti-Hebbian updates. During the Hebbian step, a training pattern is clamped to the visible layer and the resulting inter-layer phase correlations are added to \(J\). This step is calculated under a continuous $P$. We note that the fields $\psi_{A,i}$ and $\psi_{B,j}$ continue to evolve under noise during training and this stochasticity does not affect the final result. We then apply an anti-Hebbian step, where the clamp is released, both layers evolve freely, and the correlations generated during this evolution are subtracted. During the clamped phase, the hidden layer evolves for \(T_{\mathrm{clamp}}=2\), and the Hebbian correlation is taken from the fields at the end of this interval. The clamp is then removed, and both layers evolve for \(T_{\mathrm{free}}=1\) before the anti-Hebbian correlation is recorded. Both updates therefore use endpoint correlations from finite-time trajectories, rather than steady-state or time-averaged correlations. Repeating this cycle retains correlations associated with the imposed patterns while suppressing those generated by the free-running model. Polariton dynamics supplies the correlations in both steps, whereas the numerical update of \(J\) is carried out offline. Once training is complete, \(J\) is fixed; in an experimental implementation, it would be programmed into the waveguide mesh before an input is applied. The full update rules and numerical parameters are given in the SM.

For the numerical demonstration, we use ten binary templates representing the digits \(0\)--\(9\). Each template is defined on a \(5\times9\) grid, so that the visible and hidden sides each contain \(N=45\) phase variables. We stress that this task is more difficult than the traditional digit recognition task considered in polaritonics. Instead of using a supervised learning scheme, the network is not told which digit is which. Further, upon recall, the task is not just to classify an input digit, but to give the closest memorized digit in the form of $45$ output variables.

%\begin{figure}[t]
%\centering
%\includegraphics[width=\columnwidth]{Figure3.eps}
%\caption{Learning and structure of the coupling matrix.
%\textbf{a}, Initial random coupling matrix.
%\textbf{b}, Coupling matrix after Hebbian and anti-Hebbian training.
%\textbf{c}, Evolution of 15 randomly sampled coupling elements
%\(J_{ij}\) (gray) together with the matrix-averaged coupling magnitude
%\(\langle |J_{ij}|\rangle_{ij}\) (blue).
%}
%\label{Fig_Jmatrix}
%\end{figure}

\begin{figure}[b]
\centering
\includegraphics[width=\columnwidth]{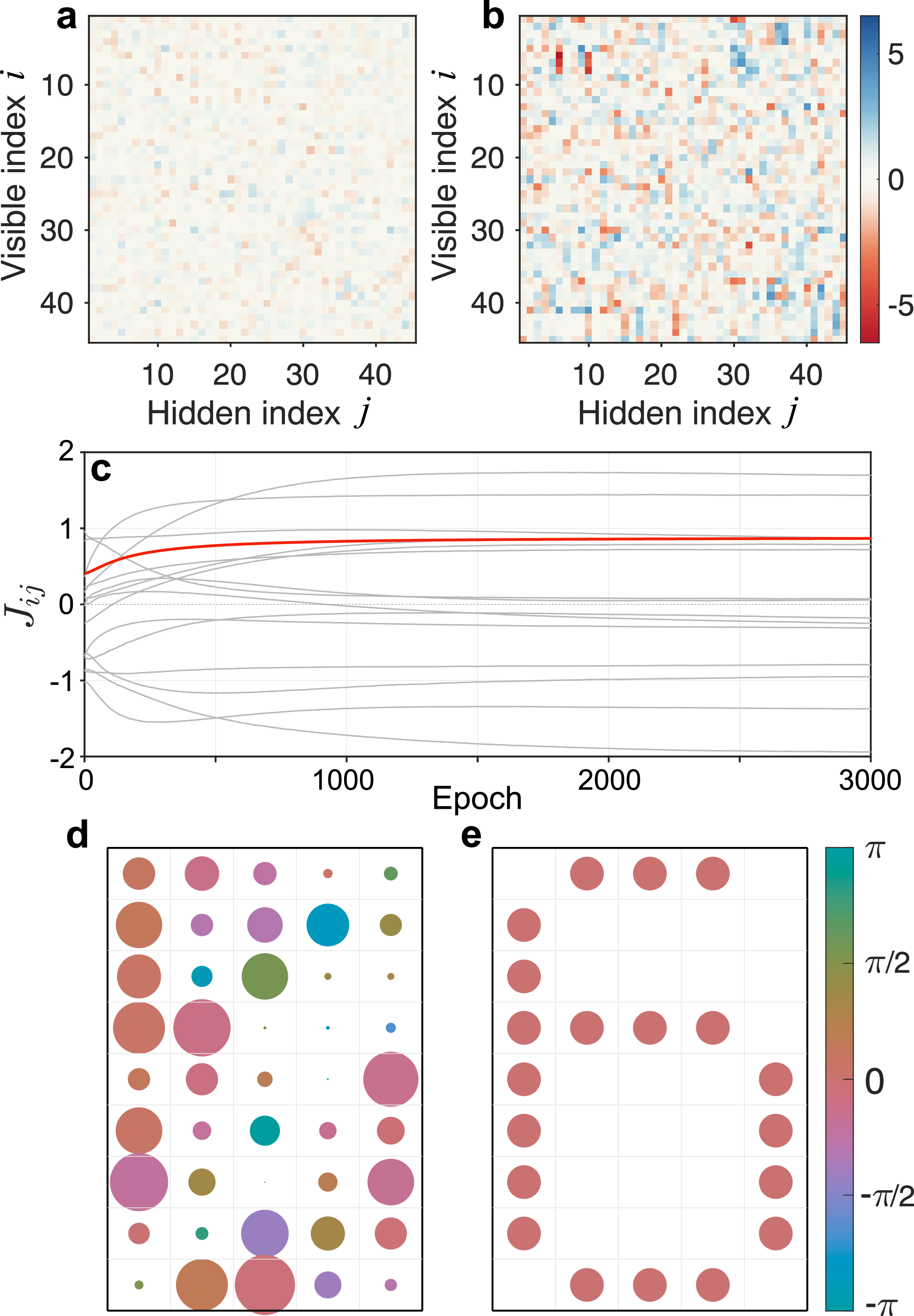}
\caption{Training of the coupling matrix and a representative
associative recall.
\textbf{a}, Initial random coupling matrix.
\textbf{b}, Coupling matrix after Hebbian and anti-Hebbian training.
\textbf{c}, Evolution of the coupling elements during training. The gray curves
show 15 randomly selected elements \(J_{ij}\), while the red curve shows the
matrix-averaged magnitude \(\langle |J_{ij}|\rangle_{ij}\).
\textbf{d}, Complex-valued corrupted input for the stored digit \(6\) at
\(\sigma_c=1\); the circle radius represents a compressed measure of
\(|z_i|^2\), and the color indicates the phase \(\arg z_i\).
\textbf{e}, Reconstructed visible-layer output obtained with the trained
coupling matrix. The corresponding target is the stored binary template of
digit \(6\).}
\label{Fig_Jmatrix}
\end{figure}

The evolution of training for the aforementioned task is shown in Fig. 3. Figs.~\ref{Fig_Jmatrix}\textbf{a} and \ref{Fig_Jmatrix}\textbf{b} show the initial (random) weight matrix and the trained result, respectively. Fig.~\ref{Fig_Jmatrix}\textbf{c} shows 15 coupling elements \(J_{ij}\) selected randomly before training, together with the matrix-averaged magnitude \(\left\langle |J_{ij}| \right\rangle_{ij}\). The sampled elements follow different, and sometimes nonmonotonic, paths, whereas the matrix-averaged magnitude increases during the early epochs and then approaches a plateau, indicating that the overall coupling scale has stabilized over the plotted training interval.

%\begin{figure}[t]
%\centering
%\includegraphics[width=0.5\textwidth]{Figure4.eps}
%\caption{
%Example of associative recall under complex-valued input corruption for
%digit \(6\). \textbf{a}, Stored target. \textbf{b}, Corrupted input at
%\(\sigma_c=1\), where the circle radius gives a common compressed
%representation of \(|z_i|^2\) and the color denotes \(\arg z_i\).
%\textbf{c}, Binarized visible-layer output. The trained
%coupling matrix is unchanged during this test.
%}
%\label{Fig_image}
%\end{figure}

\begin{figure}[t]
\centering
\includegraphics[width=0.5\textwidth]{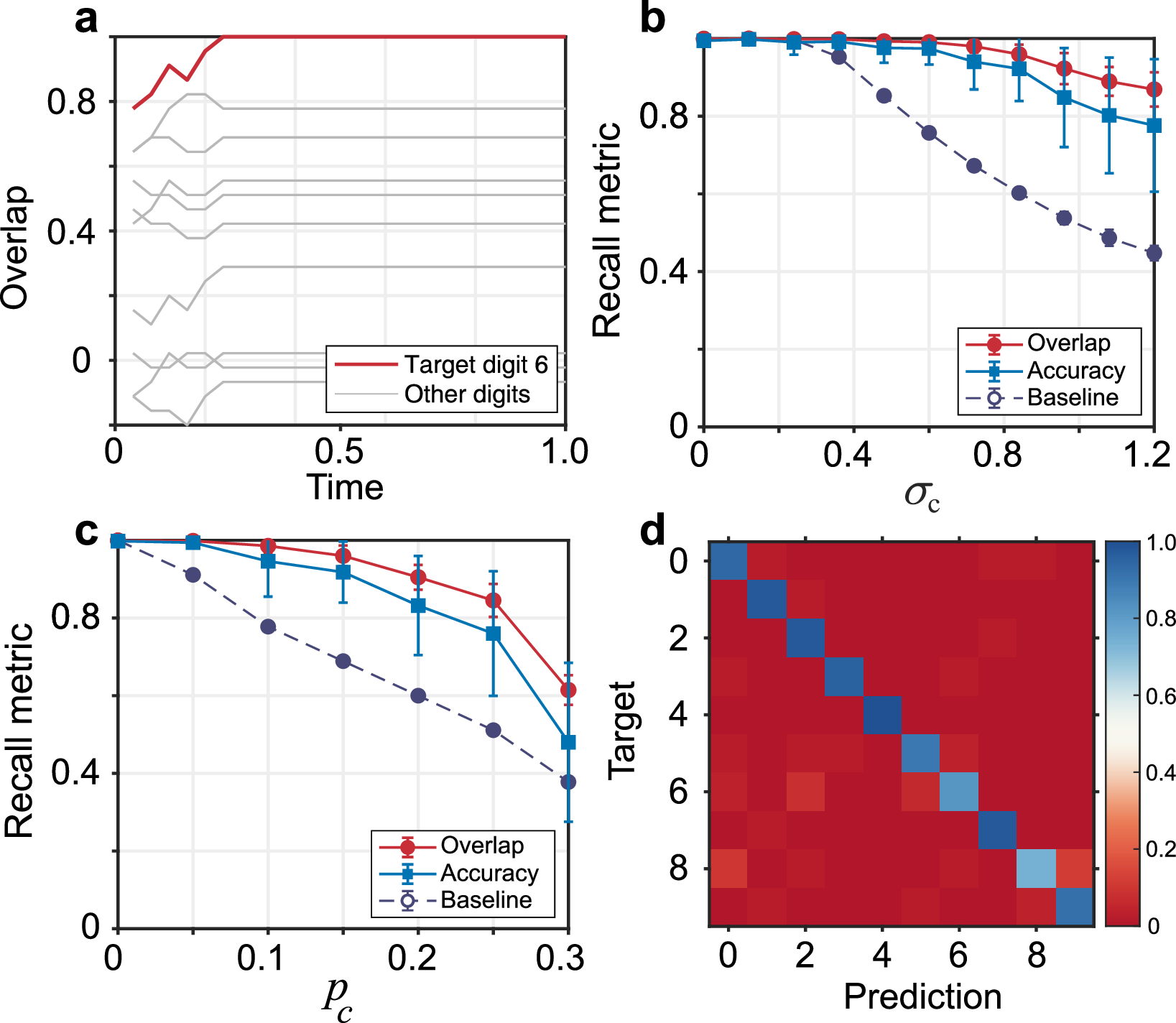}
\caption{Quantitative recall performance under complex-valued
and binary phase-flip input corruption.
\textbf{a}, Time-dependent overlap with the target digit \(6\) (red) and the
other stored digits (gray) during one recall at \(\sigma_c=0.84\).
\textbf{b}, Mean final target overlap (red circles), recovery accuracy (blue
squares), and the initial-overlap baseline (indigo dashed line) as functions of
\(\sigma_c\). The baseline is evaluated directly from the corrupted input,
before evolution under the trained coupling matrix.
\textbf{c}, The same quantities as functions of the imposed phase-flip
fraction \(p_c\).
\textbf{d}, Row-normalized confusion matrix for complex-valued corruption at
\(\sigma_c=0.84\).
For \textbf{b} and \textbf{c}, 50 independent corrupted inputs are generated for each digit at every parameter value.}
\label{Fig4}
\end{figure}

After training, the coupling matrix \(J\) is kept fixed. To properly test the network, we give it imperfect or corrupted input images and see if it can recall the appropriate memory state. We consider two forms of input corruptions here. Under binary phase-flip corruption, each pixel is independently switched between $x_i = 0$ and $x_i = 1$ with probability $p_c$, corresponding to a phase change between 0 and $\pi$. We also consider complex-valued input corruption by adding complex Gaussian noise to each binary variable,
\begin{equation}
z_i = x_i + \frac{\sigma_c}{\sqrt{2}}\left(\xi_i+i\zeta_i\right),
\end{equation}
where \(\xi_i\) and \(\zeta_i\) are independent standard normal variables for
each \(i\), and \(\sigma_c\) sets the input-corruption strength. The
corresponding visible-layer field is initialized as
\begin{equation}
\psi_{A,i}(0)=\sqrt{P-1}\left(1-2z_i\right).
\end{equation}

Figure~\ref{Fig_Jmatrix}\textbf{d} and \textbf{e} show a representative recall of digit \(6\) at \(\sigma_c=1\). Figure~\ref{Fig_Jmatrix}\textbf{d} visualizes the complex pixel variable \(z_i\): the circle radius gives a common compressed representation of \(|z_i|^2\), and the color denotes \(\arg z_i\) on a periodic scale. During recall, the hidden layer is initialized with random phases and the visible phases evolve freely under the trained inter-layer couplings. Recall results from the simultaneous evolution of all visible and hidden phases under the trained inter-layer couplings, rather than from identifying and correcting corrupted pixels one at a time. Further image-level examples for the stored digit classes under both corruption models are given in the SM.

To quantify the recall process, we calculate the overlap between the instantaneous visible-layer state and each stored digit,
\begin{equation}
O_{\mu}(t)
=
\frac{1}{N}
\sum_{i=1}^{N}
s_i(t)s_i^{(\mu)},
\label{eq:overlap}
\end{equation}
where \(s_i(t)=\operatorname{sgn}[\operatorname{Re}\psi_{A,i}(t)]\) is the binary phase state of the \(i\)th visible node and \(s_i^{(\mu)}=\cos(\pi x_i^{(\mu)})\) represents the corresponding pixel of stored digit \(\mu\). An overlap close to unity indicates that the visible-layer phases reproduce the stored pattern. Figure~\ref{Fig4}\textbf{a} follows one recall of digit \(6\) under complex-valued input corruption. Several stored patterns have appreciable overlap with the initial state, but the overlap with digit \(6\) rises above the competing values during the evolution. The final class is assigned to the stored pattern with the largest \(O_\mu(T)\). The rapid rise of \(O_6(t)\) does not imply that the continuous fields have fully relaxed. Because the overlap depends only on the signs of \(\operatorname{Re}\psi_{A,i}\), it can reach unity once the visible fields cross the appropriate binary phase boundaries, while their amplitudes and continuous phases continue to evolve as in Fig.~\ref{Fig_dynamics}.

We next evaluate all ten stored digits under the two corruption models. Figure~\ref{Fig4}\textbf{b} plots recovery accuracy and mean final target overlap against the complex-noise strength \(\sigma_c\), whereas Fig.~\ref{Fig4}\textbf{c} plots the same quantities against the imposed phase-flip fraction \(p_c\). The two corruption parameters are shown in separate panels because they describe different perturbations. Recovery accuracy is the fraction of trials for which the target digit has the largest final overlap. The target overlap instead measures the final phase agreement with the intended pattern, irrespective of the assigned class. The initial-overlap baseline, evaluated before the network evolves, decreases substantially faster than the final target overlap in both panels. Their separation shows that the trained dynamics improves agreement with the intended pattern rather than simply preserving the corrupted input. Both metrics decrease as the input corruption increases. The decrease is gradual over \(0\leq\sigma_c\leq1.2\), while the phase-flip results fall more sharply above \(p_c=0.2\). At stronger corruption, the target overlap remains above the recovery accuracy, indicating that some final states retain partial agreement with the target without being assigned to it. At \(\sigma_c=0.84\), the row-normalized confusion matrix in Fig.~\ref{Fig4}\textbf{d} remains predominantly diagonal. The remaining errors are concentrated in particular target--prediction pairs rather than spread uniformly across the digit classes. Results for partially masked digits, in which contiguous parts of the input are removed, are given in the SM.

Given the recent rise of exciton-polariton waveguides as a viable alternative for polaritonics~\citep{Souissi2026}, it is natural to revisit polaritonic architectures in ways that are compatible with their use. Here, we propose that polariton waveguide meshes offer the possibility of all-to-all coupled neural networks. We developed the theoretical foundation for a restricted Boltzmann machine architecture, identified the relevant probability distribution, and demonstrated numerically their use as an associative memory in the presence of various imperfections. To our knowledge, this represents the first scheme of a polariton-based neural network that avoids the need for supervised learning and admits unsupervised methods. Given the advances in the engineering of polariton waveguides with local phase shifts, controllable in situ, we conjecture that this architecture offers unprecedented opportunity for polaritonics in terms of flexible programmability, without pre- or post-processing, and potential compatibility with future polariton laser sources.

\textit{Acknowledgement---}
Q.X. gratefully acknowledges strong funding support from the Quantum Science and Technology-National Science and Technology Major Project (2023ZD0300300), the National Natural Science Foundation of China (grant No. 12434011) and strong support from the State Key Laboratory of Low-Dimensional Quantum Physics at Tsinghua University. T.C.H.L. was supported by the Centre for Quantum Technologies Fellowship.

\end{document}